\pdfoutput=1
\documentclass[10pt,aps,prd,twocolumn,superscriptaddress,preprintnumbers,floatfix,nofootinbib]{revtex4-2}

\usepackage{graphicx}
\usepackage{amsmath, amssymb, mathrsfs}
\usepackage{color}
\usepackage{bm}
\usepackage[T1]{fontenc}
\usepackage[utf8]{inputenc}
\usepackage{microtype}
\usepackage[normalem]{ulem}
\usepackage[dvipsnames]{xcolor}
\usepackage[colorlinks,urlcolor=NavyBlue,citecolor=NavyBlue,linkcolor=NavyBlue,pdfusetitle]{hyperref}
\usepackage[all]{hypcap}
\usepackage{orcidlink}
\usepackage{siunitx}

\graphicspath{{Scripts/figs/}{Scripts/modes-amplitude/}{Scripts/fit-comcharge/}{Scripts/vary-window/}{Scripts/window-location/}{Scripts/alphas/}{Scripts/nu-function/}}

\renewcommand{\Im}{\text{Im}}

\newcommand{\CornellPhysics}{\affiliation{Department of Physics, Cornell University, Ithaca, NY, 14853, USA}}
\newcommand{\Cornell}{\affiliation{Cornell Center for Astrophysics and Planetary Science, Cornell University, Ithaca, New York 14853, USA}}
\newcommand{\CornellLepp}{\affiliation{Laboratory for Elementary Particle Physics, Cornell University, Ithaca, New York 14853, USA}}
\newcommand{\Caltech}{\affiliation{Theoretical Astrophysics 350-17, California Institute of Technology, Pasadena, CA 91125, USA}}
\newcommand\UMiss{\affiliation{Department of Physics and Astronomy, University of Mississippi, University, MS 38677, USA}}

\newcommand{\pd}{\partial}

\def\t{\theta}
\def\p{\phi}
\def\f{\frac}
\def\nn{\nonumber}

\newcommand{\figAmplitudes}{%
\begin{figure}
\centering
\includegraphics[width=\linewidth]{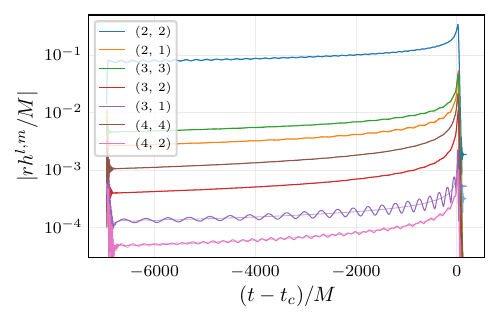}
\vspace{-1.5em}
\label{fig:amplitudefig}
\caption{Oscillatory pattern in the amplitude of different $(\ell,m)$ modes due to
waveform being in an arbitrary BMS frame for a quasicircular nonprecessing system
\texttt{SXS:BBH:2115}.
The time axis has been shifted by the common horizon time, $t_{c}$.
The dark curves represent the mode amplitudes when the
waveform is in an arbitrary BMS frame, while the light curves are obtained after
fixing the BMS frame of the waveform. The power from the dominant $(2,\pm 2)$
modes leaks to subdominant higher order modes. Fixing the frame eliminates the
oscillations.}
\end{figure}%
}

\newcommand{\figComCharge}{%
\begin{figure}
\centering
\includegraphics[width=\linewidth]{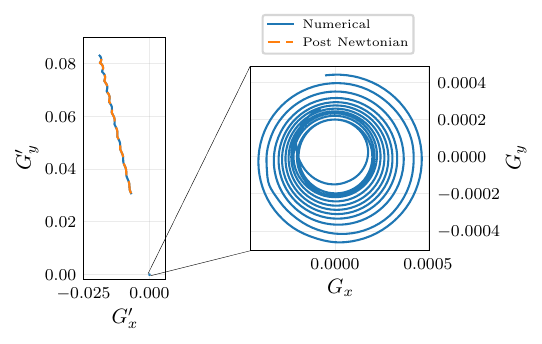}
\vspace{-1.5em}
\label{fig:com charge}
\caption{%
The center-of-mass charge vector $\vec{G}$ (in blue) obtained from the
asymptotic data of a quasicircular nonprecessing system \texttt{SXS:BBH:2115}.
Before frame-fixing (left panel), $\vec{G}$ starts closer to the origin and drifts away in a boosted
outspiral. A fit to the boosted post-Newtonian prediction
of Eq.~\eqref{eq:Boosted CoM charge} (orange, dashed)
agrees well with numerical data.
Notice the
difference in scales after fixing the PNBMS frame (right panel). The small box
near the origin in the left plot correspond to the range of CoM charge after
fixing the frame.
This plot includes times in the 15\%--85\% region between
the metadata's
\texttt{reference\_time} and
the time of the maximum norm of $h$ across the entire 2-sphere.
}
\end{figure}%
}

\newcommand{\tabSimulations}{%
\begin{table}[tb]
\centering
\begin{tabular}{c|SSS}
  \noalign{\smallskip} \hline \noalign{\smallskip} \hline \noalign{\smallskip}
 {SXS ID} & {$q$} & {$\chi^1_z$} & {$\chi^2_z$} \\
 \noalign{\smallskip} \hline \noalign{\smallskip}
SXS:BBH:3928 & 1.28 & 0.3 & -0.1 \\
SXS:BBH:2331 & 1.5 & -0.0 & -0.0 \\
SXS:BBH:2337 & 1.5 & -0.5 & 0.0 \\
SXS:BBH:2115 & 2.0 & -0.3 & 0.0 \\
SXS:BBH:2120 & 2.0 & 0.0 & 0.3 \\
SXS:BBH:2124 & 2.0 & 0.3 & -0.0 \\
SXS:BBH:2143 & 3.0 & -0.3 & -0.0 \\
SXS:BBH:2154 & 3.0 & 0.3 & -0.0 \\
SXS:BBH:1221 & 3.0 & 0.5 & -0.0 \\
SXS:BBH:1911 & 4.0 & -0.0 & -0.8 \\
SXS:BBH:1942 & 4.0 & 0.4  & -0.8 \\
SXS:BBH:2013 & 4.0 & 0.0 & 0.4 \\
SXS:BBH:2374 & 5.0 & -0.0 & 0.0 \\
SXS:BBH:3619 & 5.0 & 0.0 & 0.0 \\
SXS:BBH:2168 & 6.0 & -0.0 & -0.8 \\
SXS:BBH:2225 & 6.0 & -0.8 & 0.4 \\
SXS:BBH:1429 & 7.75 & -0.2 & -0.8 \\
SXS:BBH:2677 & 8.0 & 0.4 & 0.8 \\
SXS:BBH:2696 & 8.0 & 0.8 & -0.8 \\
SXS:BBH:4235 & 10.0 & -0.8 & 0.0 \\
\noalign{\smallskip} \hline \noalign{\smallskip} \hline \noalign{\smallskip}
\end{tabular}\label{tab:simulations}
\caption{%
  Parameters of all BBH simulations used in the analysis. $q = M_1/M_2$
  is the mass ratio (rounded to two decimal places), and $(\chi_1^z,\chi_2^z)$
  are the $z$ component of dimensionless spins of the two black holes (rounded
  to one decimal place).}
\end{table}%
}

\newcommand{\figWindowChoices}{%
\begin{figure}
\centering
\includegraphics[width=\linewidth]{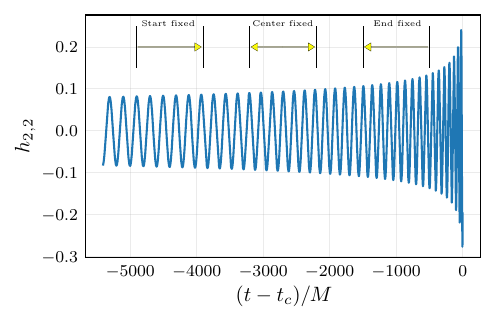}
\vspace{-2em}
\label{fig:window-choices}
\caption{Different choices for varying the window sizes over the inspiral. The
arrow head (in yellow) represents the direction in which the window size is
increased for our analysis.}
\end{figure}%
}

\newcommand{\figCombinedBoostTransComp}{%
\begin{figure*}[t]
\centering
\includegraphics[width=\linewidth]{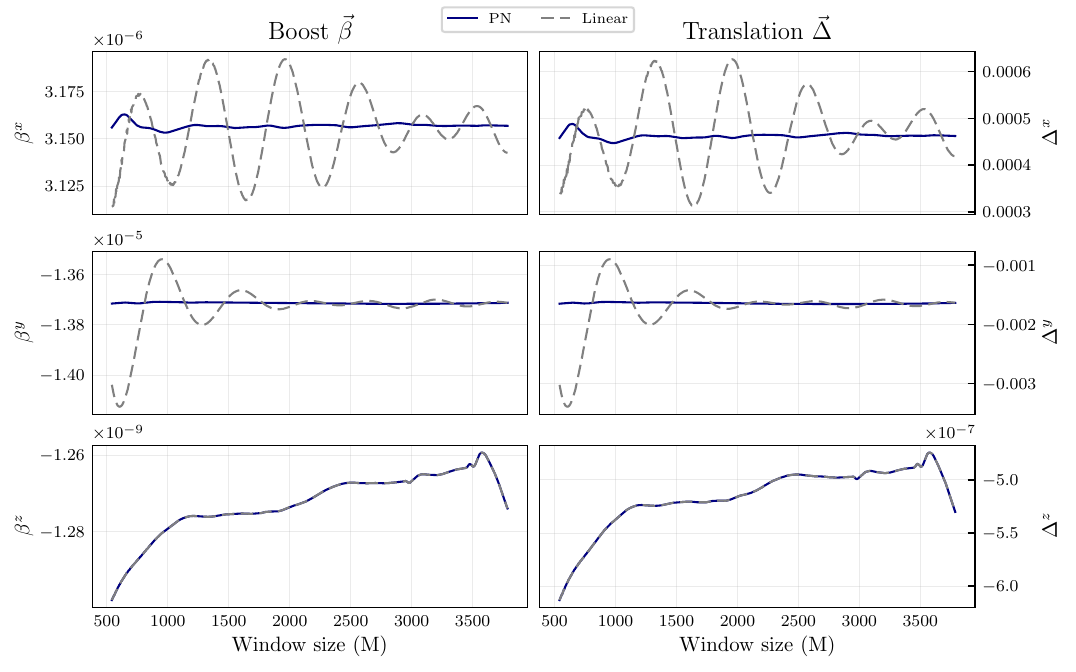}
\vspace{-1.5em}
\label{fig:sensitivity plot}
\caption{Sensitivity of the boost $\vec{\beta}$ and translation $\vec{\Delta}$
fit parameters to different window sizes for fitting the boosted CoM charge
for the simulation \texttt{SXS:BBH:2115}, which shows some of the best
improvement between the old and new methods.
For example, by the metric of $\sigma^2_{\text{linear}}/\sigma^2_{\text{PN}}$,
the new fit for $\beta^{y}$ is less variable by a factor of
1580, and the new fit for $\Delta^{y}$ is less variable by a factor of 1400
(see Table~\ref{tab:fit-median-improvement} for typical values).
The dashed grey curve shows the sensitivity of
the previous linear fit while the navy blue curve shows the sensitivity of the new
analytical PN fit from Eq.~\eqref{eq:Boosted CoM charge}.
The windows for this figure were ``center fixed'' (see
Fig.~\ref{fig:window-choices}), but is qualitatively similar for start- and
end-fixed.
}
\end{figure*}%
}

\newcommand{\figfvsnu}{%
\begin{figure}
\includegraphics{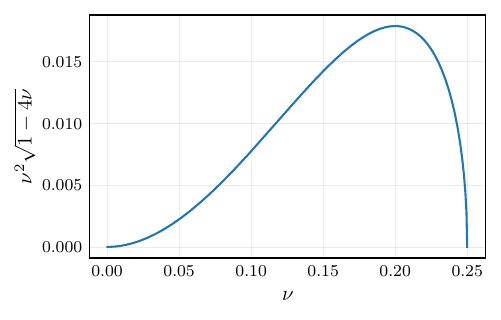}
\caption{%
  Prefactor in the analytical CoM charge expression with derivative undefined at
  $\nu = \tfrac{1}{4}$, which is the case for equal mass ratio binaries.}
\label{fig:fvsnuplot}
\end{figure}%
}

\newcommand{\tabfitmedianimprovement}{%
\begin{table}[tb]
\centering
\begin{tabular}{c|SSSS}
\noalign{\smallskip} \hline \noalign{\smallskip} \hline \noalign{\smallskip}
med $\left( \dfrac{\sigma^2_{\text{linear}}}{\sigma^2_{\text{PN}}} \right)$   & {$\beta^{x}$} & {$\beta^{y}$} & {$\Delta^{x}$} & {$\Delta^{y}$} \\
\noalign{\smallskip} \hline \noalign{\smallskip}
End fixed  & 9.7 & 1.9 & 10.4 & 2.0 \\
Center fixed  & 24.8 & 17.6 & 20.0 & 11.1 \\
Start fixed & 1.7 & 1.1 & 1.7 & 1.1 \\
\noalign{\smallskip} \hline \noalign{\smallskip} \hline \noalign{\smallskip}
\end{tabular}
\caption{%
  \label{tab:fit-median-improvement}
  Median (across all simulations) of ratio of variances for each fit parameter,
  showing improved robustness of the PN-based fit. For each simulation, the fit
  parameters are obtained using the linear and PN based fit for the center of
  mass charge over the numerical data. We obtain the fit parameters for
  different window sizes ranging from approximately 500M to 4000M. We then
  evaluate the variance for each parameter from this set, and hence the ratio of
  variances for the two fit choices (linear and PN). We repeat this comparison
  for the 3 choices of variation of window size.
}
\end{table}
}

\newcommand{\figAlphas}{%
\begin{figure}
\centering
\includegraphics[width=\linewidth]{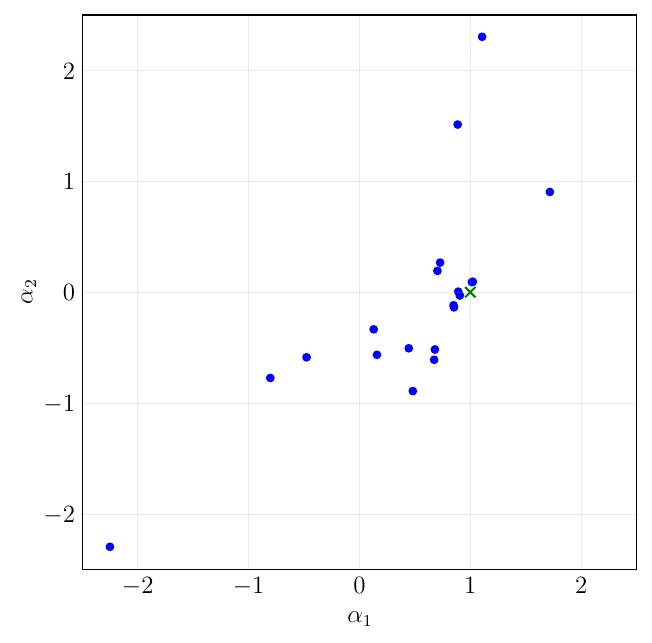}
\vspace{-2em}
\label{fig:alphas}
\caption{%
  The distribution of nuisance parameters $(\alpha_1, \alpha_2)$ (represented by
  blue dots, one for each simulation) with a fixed window size across all the
  simulations used in our analysis. The size of the window is fixed to be 1500M,
  and placed at the center of the inspiral.  The green cross denotes the leading-order PN
  prediction.}
\end{figure}%
}

\begin{document}

\title{Fixing the center-of-mass frame of numerical relativity waveforms using the post-Newtonian center-of-mass charge}

\author{Aniket Khairnar\,\orcidlink{0000-0001-5138-572X}}
\email{akhairna@go.olemiss.edu}
\UMiss

\author{Leo C.\ Stein\,\orcidlink{0000-0001-7559-9597}}
\UMiss

\author{\\ Michael Boyle\,\orcidlink{0000-0002-5075-5116}}
\Cornell

\author{Nils Deppe\,\orcidlink{0000-0003-4557-4115}} \CornellLepp \CornellPhysics \Cornell
\author{Lawrence E.~Kidder\,\orcidlink{0000-0001-5392-7342}} \Cornell
\author{Keefe Mitman\,\orcidlink{0000-0003-0276-3856}}
\Cornell
\author{Jordan Moxon\,\orcidlink{0000-0001-9891-8677}} \Caltech
\author{Kyle C.~Nelli\,\orcidlink{0000-0003-2426-8768}} \Caltech
\author{William Throwe\,\orcidlink{0000-0001-5059-4378}} \Cornell
\author{Nils L.~Vu\,\orcidlink{0000-0002-5767-3949}} \Caltech

\hypersetup{pdfauthor={Khairnar, Stein, et al.}}

\date{\today}

\begin{abstract}
The Bondi--van der Burg--Metzner--Sachs (BMS) frame of gravitational waves
produced by numerical relativity (NR) simulations is crucial for building
accurate waveform models. A proper comparison of NR waveforms with other
models requires fixing the arbitrary BMS frame. In this work we
improve the center-of-mass (CoM) frame fixing for quasicircular, nonprecessing
binary systems.
Past work approximated the CoM motion with just a linear fit. We
compute a post-Newtonian result of the boosted CoM charge to also capture its
physical out-spiraling oscillations.
We show that using the analytical results improves the
robustness of the fit parameters---translation and boost vectors---to the choice
of duration and time of the fitting window. Our analysis
demonstrates a maximum improvement in robustness when the window is placed at
the center of the inspiral. We quantified this
improvement by computing the ratio of variances of fit parameters when the fit
window size is varied. The largest improvement in robustness of
parameters is by a factor of $\sim 25$ for the boost vector and $\sim 20$ for
the translation vector. Finally, we incorporate this method
into the BMS frame-fixing routine of the python package \texttt{scri} for
waveforms produced with Cauchy-characteristic evolution.
\end{abstract}

\maketitle

\section{Introduction}
\label{sec:introduction}

The LIGO-Virgo-Kagra (LVK) collaboration have detected hundreds of gravitational
wave (GW) events since 2015~\cite{LIGOScientific:2018mvr,LIGOScientific:2020ibl,KAGRA:2021vkt}. The
sensitivity of the detectors has significantly improved since the first
observation run~\cite{KAGRA:2013rdx}, and continues to improve in the future
observation era. In addition, we expect
future ground based detectors (Cosmic Explorer, Einstein Telescope), and space
based detector LISA~\cite{LISA:2017pwj} to join the gravitational wave network
each with its own unique location in the frequency band.

All these improvements and new detectors will present data that can help us in
testing our theory of gravity in the strong field regime. Our GW
detection and parameter estimation capability also depends on the accuracy of our
waveform models. It is an impossible task to handle the full complexity of the
Einstein's field equations analytically. Therefore, we rely on numerical
methods to solve the field equations and generate solutions.

The most accurate models for gravitational waves are the waveforms produced
by numerical relativity simulations~\cite{Mroue:2013xna,Boyle:2019kee,Scheel:2025jct}.
Numerical relativity solves Einstein's
equations on supercomputers with a high degree of accuracy. However,
advances on the experimental side of gravitational wave science demand
further improvements in NR. This involves exploring challenging
regions of parameter space, improving truncation errors of NR simulations,
producing simulations with matter fields, building surrogate models, improving
wave extraction procedures, and understanding gauge effects.

Apart from the modeling challenges mentioned above, we know that NR waveforms
are finite and computationally expensive. The data
analysis of GWs requires a comprehensive template bank consisting of a large
collection $(\sim 10^6)$ of highly accurate and computationally efficient
waveforms. Typically, such template banks are constructed using
phenomenological (Phenom) models~\cite{Pratten:2020fqn, Bernuzzi:2011aq},
the effective-one-body (EOB)
method~\cite{Hinderer:2013uwa, Ramos-Buades:2023ehm, Gamba:2024cvy, Khalil:2023kep},
and surrogate models~\cite{Varma:2018mmi,Varma:2019csw,Yoo:2022erv}.
Either these models use NR waveforms for calibration during merger and
ringdown (like in Phenom and EOB models) or are trained purely on NR waveforms
to accurately model the signal. Therefore, the accuracy of such models are
bounded by the accuracy of NR waveforms that were used for calibration and training.

An often overlooked effect in waveform modeling is the choice of frame that the
gravitational waveforms are in. For NR simulations the asymptotic waveforms
extracted at asymptotic null infinity ($\mathscr{I}^+$) are
certainly not gauge invariant, rather they are dependent on the choice of
coordinate system. The choice of gauge conditions in the numerical setup
produces waveforms in an arbitrary numerical frame, and can
lead to finite amplitude gauge differences in waveforms~\cite{Boyle:2015nqa}.
Thus, there is a need to minimize these artificial effects by systematically
fixing the frame of asymptotic waveforms. The asymptotic frame for an isolated
gravitational system is best described by using the asymptotic symmetry group at
null infinity which is not the usual Poincaré group, but an extension of it
called the Bondi--van der Burg--Metzner--Sachs (BMS)
group~\cite{Bondi:1962px,Sachs:1962wk,Sachs:1962zza,Bondi:1960jsa, Sachs:1961zz}
that includes supertranslations.

Previous work~\cite{Boyle:2015nqa} has examined the transformation of
gravitational waveform modes under the BMS group. The process of
transforming a waveform from one BMS frame to another can generate appreciable
differences in waveform modes even when the transformations are small.
Therefore, waveform models must be expressed in the same frame to prevent
gauge choices from introducing errors during comparisons. This motivates
the search for a desirable BMS frame, and transformations to fix the
gauge freedom of gravitational waveforms.

The gauge freedom of SXS waveforms was first fixed using a center-of-mass
(CoM)
correction procedure 
developed in~\cite{Woodford:2019tlo,Boyle:2015nqa,Boyle:2019kee}. The procedure
used the coordinate trajectories of the apparent horizons obtained from the simulations to
attain the boost and translation parameters required for fixing the center of
mass frame. However, this method fixed only six
degrees of freedom and there exist infinite degrees of freedom remaining in
the BMS group to be fixed. The gauge fixing procedure was also limited by the
amount of information available from the output of the extraction method and
fixing the entire BMS gauge was left for future work.

Gauge effects arise in all methods of
gravitational wave extraction to
$\mathscr{I}^+$. References~\cite{Moxon:2021gbv, Moxon:2020gha, Iozzo:2020jcu}
developed the SpECTRE Cauchy-characteristic evolution (CCE) scheme to supersede
the older approach of polynomial extrapolation of gravitational waves from SXS
simulations. The CCE method evolves not just the
strain, but the full set of Weyl scalars out to $\mathscr{I}^{+}$. For CCE waveforms one can
fix the entire BMS freedom instead of just the center-of-mass frame using the
information available from Weyl
scalars. References~\cite{Mitman:2021xkq, Mitman:2022kwt} developed the approach
of BMS frame-fixing by computing
Bondi charges from asymptotic data, i.e., the strain and Weyl scalars.
The desired frame is achieved by finding the BMS transformation
that transforms the charges to some target values. Thus, the NR waveform is mapped
to the PNBMS frame or superrest frame (both defined below) using knowledge of
the BMS charges in these respective frames.

The older center-of-mass frame fixing approach can be inaccurate and is sensitive to the
time-window one chooses when performing BMS frame fixing.
We propose an improvement over this method by using analytical knowledge of
the CoM charge which we obtain from PN theory.
We use this PN result to derive the boosted center-of-mass charge that can be
used for fitting the boost and translation vectors. Our improved method improves
the robustness of the fit to the choice of window duration and its location.
In this work we restrict ourselves to quasicircular nonprecessing systems, as
the relevant PN results are not available for generic systems. This work is a
precursor to fixing the entire BMS gauge for precessing as well as eccentric systems
which we leave for future work.

We start by discussing the effects of BMS gauge transformations on waveform modes
in Section~\ref{sec: effect of BMS}.
Section~\ref{sec:gauge-effects} demonstrates the effect of gauge choice on NR
waveforms.
Section \ref{sec: frame fixing
using charges} explains the BMS frame-fixing procedure (including the CoM freedom)
using charges, for both the PNBMS and superrest frames.
Section~\ref{sec: frame fixing using PN} derives the PN prediction for the boost
charge which we use to perform the frame fixing operation, and
Section~\ref{sec:sensitivity-analysis} performs a sensitivity analysis to
demonstrate how the new method performs better than the previous approach.
We end with conclusions and future work in Section \ref{sec:disc-concl}.

\subsection{Conventions}
We follow the conventions for the derivative operators $\eth$ and $\bar{\eth}$ from
previous works \cite{Iozzo:2021vnq,Mitman:2021xkq,Mitman:2022kwt}, which also
include those for gravitational wave strain $h$, the shear $\sigma$, and the Weyl scalars
$\Psi_{0 - 4}$. For a function $f(u,\theta,\phi)$ with spin-weight $s$,
the actions of $\eth$ and $\bar{\eth}$ are
\begin{subequations}
	\begin{align}
	\eth f(u,\theta,\phi)&=-\frac{1}{\sqrt{2}}(\sin\theta)^{+s}(\partial_{\theta}+i\csc\theta\partial_{\phi})\nonumber\\
	&\phantom{=.-\frac{1}{\sqrt{2}}(\sin(\theta))^{s}}\left[(\sin\theta)^{-s}f(u, \theta,\phi)\right],\\
	\bar{\eth} f(u,\theta,\phi)&=-\frac{1}{\sqrt{2}}(\sin\theta)^{-s}(\partial_{\theta}-i\csc\theta\partial_{\phi})\nonumber\\
	&\phantom{=.-\frac{1}{\sqrt{2}}(\sin(\theta))^{s}}\left[(\sin\theta)^{+s}f(u,\theta,\phi)\right].
	\end{align}
\end{subequations}
We decompose any spin-$s$ function using spin-weighted spherical harmonics as
\begin{equation}
	f(u, \theta, \phi) = \sum\limits_{\ell,m}f_{\ell, m}(u)\,{}_{s}Y_{\ell, m}(\theta,\phi).
\end{equation}
Thus, when acting on spin-weighted spherical harmonics, the operators produce
\begin{subequations}
	\begin{align}
	\eth({}_{s}Y_{\ell, m})&=+\frac{1}{\sqrt{2}}\sqrt{(\ell-s)(\ell+s+1)} \, {}_{s+1}Y_{\ell, m},\\
	\bar{\eth}({}_{s}Y_{\ell, m})&=-\frac{1}{\sqrt{2}}\sqrt{(\ell+s)(\ell-s+1)} \, {}_{s-1}Y_{\ell, m}.
	\end{align}
\end{subequations}

\section{BMS group and the transformation of waveform}
\label{sec: effect of BMS}
In asymptotically flat spacetimes, it is possible to find a collection of
coordinates adapted to the universal asymptotic structure.  Such coordinates
orginated through the seminal works of Bondi, van der Burg, Metzner, and Sachs
in the 1960s~\cite{Bondi:1962px, Bondi:1960jsa, Sachs:1961zz, Sachs:1962wk,
  Sachs:1962zza}.  These are known as Bondi-Sachs coordinates, which consist of
a retarded time coordinate $u$, a radial coordinate $r$, and angular coordinates
$(\theta, \phi)$. The metric for asymptotically flat spacetimes can be expressed
in Bondi-Sachs coordinates as~\cite{Flanagan:2015pxa}
\begin{align}
	ds^2 = &- U e^{2 \beta} d u^2 + 2 e^{2\beta} du dr  \nonumber \\ 
	& + r^2 \gamma_{AB} (dx^A - \mathcal{U}^A du)(dx^B - \mathcal{U}^B du), \label{eq: Bondi-metric}
\end{align}
where the capital Latin indices range over the angular coordinates. $U$,
$\beta$, $\mathcal{U}^A$, and $\gamma_{AB}$ are functions of the Bondi
coordinates $(u, r, \theta, \phi)$.
In order for the metric to be asymptotically flat, the metric functions should
have the asymptotic behavior~\cite{Flanagan:2015pxa}
\begin{subequations}
\label{eq:Bondi-gauge}
\begin{align}
U&\rightarrow1,\\
\beta&\rightarrow0,\\
\mathcal{U}^{A}&\rightarrow0,\\
\label{eq:Bondi-gauge-angular}
\gamma_{AB}&\rightarrow\begin{pmatrix}1&0\\0&\sin^{2}\theta\end{pmatrix},
\end{align}
\end{subequations}
as $r \rightarrow \infty$.
To reduce the full diffeomorphism freedom down to only the asymptotic symmetry
group, one imposes an additional gauge condition of $\partial_r
\text{det}(\gamma_{AB}) = 0$. This condition, together with the Bondi-Sachs
coordinate conditions $g_{rr}=0$ and $g_{rA}=0$, are
referred to as the ``Bondi gauge.'' The Bondi metric along with the gauge
conditions represents a class of metrics describing isolated,
asymptotically-flat gravitational systems. It
doesn't account for the entire gauge freedom of general relativity. There are
infinitely many gauge transformations preserving the structure of the metric in
Eq.~\eqref{eq: Bondi-metric} and the fall-off behavior in Eq.~\eqref{eq:Bondi-gauge}.
These gauge transformations form the asymptotic symmetry group known as the BMS
group. The BMS group extends the Poincaré group with spacetime translations
replaced by angle-dependent translations called supertranslations.

Our universe does not seem to be asymptotically flat, and real
gravitational-wave detectors are at finite distances from sources. Nevertheless,
we will still make use of the structure of asymptotically-flat spacetimes.  We
can study gravitational waveforms using Bondi gauge in the vicinity of future
null infinity $\mathscr{I}^+$, which is the conformally compactified boundary of
spacetime in the limit $r \rightarrow \infty$ along outgoing null rays.  It is
important to understand
how gravitational waveforms transform under the action of an element of the BMS
group.  The transformation of coordinates and the asymptotic data under BMS
transformations was investigated in~\cite{Boyle:2015nqa}.  These transformations
are simpler in terms of $u$ and a complex stereographic
coordinate $\zeta$ on the sphere defined as
\begin{align}
\zeta \equiv e^{i\phi} \cot\left(\frac{\theta}{2}\right).
\end{align}
A BMS transformation acts on $(u,\zeta)$ as~\cite{Boyle:2015nqa,Moreschi:1998mw}
\begin{align}	
\label{eq:coordtransformation}
\left(u, \zeta\right) \rightarrow \left(u', \zeta'\right) = \left((u - \alpha) \, k(\zeta, \bar{\zeta}), \frac{a\zeta + b}{c\zeta + d}\right),
\end{align}
where the conformal factor $k(\zeta, \bar{\zeta})$ is given by
\begin{align}
\label{eq:conformalfactor}
k(\zeta, \bar{\zeta}) \equiv \frac{1 + \zeta \bar{\zeta}}{(a\zeta + b)(\bar{a} \bar{\zeta} + \bar{b}) + (c\zeta + d)(\bar{c} \bar{\zeta} + \bar{d})}.
\end{align}
The parameters $(a,b,c,d)$ are complex coefficients satisfying $ad - bc = 1$, which encode the Lorentz transformations. The function $\alpha(\zeta, \bar{\zeta})$ is a real-valued, smooth function on the celestial sphere and parameterizes a supertranslation.

Under the above BMS transformation, the asymptotic fields --- the Bondi shear
$\sigma$ and the Weyl scalars $\Psi_A$ --- transform as~\cite{Boyle:2015nqa}
\begin{subequations}
\label{eq:vartransformation}
\begin{align}
\label{eq:sheartransformation}
\sigma' &= \frac{e^{2i\lambda}}{k} \left[ \sigma - \eth^2 \alpha \right], \\
\label{eq:psitransformation}
\Psi_A' &= \frac{e^{(2 - A)i\lambda}}{k^3} \sum_{a = A}^{4} \begin{pmatrix} 4 - A \\ a - A \end{pmatrix} \left( -\frac{\eth u'}{k} \right)^{a - A} \Psi_a,
\end{align}
\end{subequations}
where $A \in \{0, 1, 2, 3, 4\}$, and $\lambda$ is the spin phase ~\cite{Boyle:2015nqa}, defined as
\begin{align}
\exp(i\lambda) = \left[\frac{\partial \bar{\zeta}'}{\partial \bar{\zeta}} \left( \frac{\partial \zeta'}{\partial \zeta} \right)^{-1} \right]^{1/2} = \frac{c\zeta + d}{\bar{c} \bar{\zeta} + \bar{d}}.
\end{align}
The transformations in Eqs.~\eqref{eq:coordtransformation}
and~\eqref{eq:vartransformation} will take place when performing BMS
frame-fixing, including CoM fixing, as described below.

\section{Gauge effects in waveform}
\label{sec:gauge-effects}
As mentioned earlier the choice of gauge can affect a numerical relativity
simulation. The initial data of the simulations represent a ``snapshot'' of the
spacetime at the beginning of the simulation. This snapshot does not accurately
capture the entire past history of the binary before the start of the
simulation, and must make a choice for the free data.
Therefore, we expect differences in waveforms due to different
choices made while constructing initial data.

References~\cite{Boyle:2015nqa, Woodford:2019tlo,Ossokine:2015yla} identified
large displacement and drift in the position of the center of mass from raw data
in simulations in the SXS catalog---a gauge effect present in
all simulations. This effect is independent of the physical
parameters of the system and resolution.
The displacement and motion of the center of mass cause mode
mixing in gravitational waveforms, and thus power from the dominant
$(2, \pm 2)$ modes mixes into the less dominant, higher-order modes. This can
be seen as amplitude modulation in the higher-order modes.
\figAmplitudes
Fig.~\ref{fig:amplitudefig} shows these oscillations in the dominant and
subdominant modes for the quasicircular, nonprecessing system
\texttt{SXS:BBH:2115}.

References~\cite{Boyle:2015nqa, Woodford:2019tlo} introduced a
center-of-mass correction procedure to fix the boost and translation
freedom of extrapolated waveforms. This procedure used Newtonian trajectories of
the apparent
horizons to attain the trajectory of the center of mass of the binary. This
was then used to find the boost and translations that
would map the system to the center-of-mass frame through a minimization
procedure. However, this method is limited by the use of coordinate-dependent
quantities that are defined in the bulk of spacetime, and by the leading order
Newtonian results for the center of mass.

A better assessment of the center-of-mass location is through the CoM charge
that can be obtained purely from the asymptotic data. This charge, $\vec{G}(u)$,
was computed numerically in reference \cite{Mitman:2021xkq} for the purpose of
fixing the CoM frame for CCE waveforms.
This method is only available for CCE waveforms, since they include
the Weyl scalars needed to compute $\vec{G}$, whereas the older extrapolation
procedure lacks the scalars $\Psi_{0-3}$.
Figure~\ref{fig:com charge} visualizes $\vec{G}(u)$ obtained from a
quasicircular nonprecessing NR simulation. It clearly
exhibits a linear drift as well as oscillations. The linear drift is due to
imperfect initial data, wherein the system starts out boosted with respect to
the center-of-mass frame. The oscillations in the CoM charge are physical:
because of conservation of linear momentum, the center of mass makes an
out-spiral in time.
We show the behavior of the center-of-mass charge after fixing the frame in the
right panel.
Notice the difference in
scales of the two panels. The tiny box near the origin in the left panel
corresponds to the range of motion of the center of mass after fixing the frame.

\figComCharge

In addition to fixing the center-of-mass frame it is desirable to fix the other
BMS degrees of freedom like rotations and $\ell \geq 2$ supertranslations. Using
the asymptotic Bondi data, it is possible to fix the entire BMS gauge freedom.
For example, we can derive a rotation ``charge'' and the supertranslation charge from
the strain and Weyl scalars, and map the system to their desired values in the
PNBMS frame or the superrest frame. In the next section we review
how these charges are obtained from the asymptotic data and then used for fixing
the frame.

\section{Frame fixing using BMS charges}
\label{sec: frame fixing using charges}
Reference~\cite{Mitman:2022kwt} introduced a procedure for fixing a BMS frame by
making use of BMS charges computed from
asymptotic Bondi data. These charges include the four-momentum charge $P^a$, the
angular momentum charge $\vec{J}$, the
boost charge $\vec{K}$, the center-of-mass charge $\vec{G}$, and the
supertranslation charge $\Psi_{M}$. The charges are computed by
taking moments of the Bondi mass aspect $m$, the Lorentz aspect $N$, and the energy
moment aspect $E$ on the celestial two-sphere. The aspects are obtained from
the expansion of the Bondi-Sachs metric in Bondi gauge. In the MB\footnote{Moreschi-Boyle conventions used in the
works~\cite{Boyle:2015nqa,Iozzo:2020jcu,Boyle:2013nka,Moreschi:1998mw,Moreschi:1988pc,Iozzo:2021vnq} and the code
\texttt{scri}~\cite{boyle_2025_15693419}. The complex dyads used for building
the spin-weighted fields are discussed in detail in these references.}
conventions they are related to the strain and Weyl scalars as
\begin{subequations}
\begin{align}
m(u,\theta,\phi)&\equiv-\text{Re}\left[\Psi_{2}+\sigma\dot{\bar{\sigma}}\right],\\
N(u,\theta,\phi)&\equiv-\left(\Psi_{1}+\sigma\eth\bar{\sigma}+u\eth 
m+\frac{1}{2}\eth\left(\sigma\bar{\sigma}\right)\right),\\
E(u,\theta,\phi)&\equiv N+u\eth m\nonumber\\
&=-\left(\Psi_{1}+\sigma\eth\bar{\sigma}+\frac{1}{2}\eth\left(\sigma\bar{\sigma}\right)\right).
\end{align}
\end{subequations}
From these aspects we obtain the Poincaré charges as
\begin{subequations}
\label{eq:Poincarécharges}
\begin{align}
\label{eq:momentumcharge}
P^{a}(u)&=\frac{1}{4\pi}\int_{S^{2}}n^{a} m\,d\Omega,\\
\label{eq:rotationcharge}
J^{a}(u)&=\frac{1}{4\pi}\int_{S^{2}}\text{Re}\left[\left(\bar{\eth}n^{a}\right)\left(-iN\right)\right]\,d\Omega,\\
\label{eq:boostcharge}
K^{a}(u)&=\frac{1}{4\pi}\int_{S^{2}}\text{Re}\left[\left(\bar{\eth}n^{a}\right)N\right]\,d\Omega,\\
\label{eq:CoMcharge}
G^{a}(u)&=\left(K^{a}+uP^{a}\right)/P^{0}\nonumber\\
&=\frac{1}{4\pi}\int_{S^{2}}\text{Re}\left[\left(\bar{\eth}n^{a}\right)\left(N+u\eth 
m\right)\right]\,d\Omega/P^{0},
\end{align}
\end{subequations}
where $n^{a}$  is a collection of spin-0 scalar functions. The scalar components
are combinations of the $\ell\leq 1$ spherical harmonics,
\begin{subequations}
\begin{align}
n^{t}&=1 \nonumber\\
&=\sqrt{4\pi}Y_{0,0},\\
n^{x}&=\sin\theta\cos\phi, \nonumber\\
&=\sqrt{\frac{4\pi}{3}}\left[\frac{1}{\sqrt{2}}\left(Y_{1,-1}-Y_{1,+1}\right)\right],\\
n^{y}&=\sin\theta\sin\phi \nonumber\\
&=\sqrt{\frac{4\pi}{3}}\left[\frac{i}{\sqrt{2}}\left(Y_{1,-1}+Y_{1,+1}\right)\right],\\
n^{z}&=\cos\theta \nonumber\\
&=\sqrt{\frac{4\pi}{3}}Y_{1,0},
\end{align}
\end{subequations}

Note that our definition of CoM charge ($G^a$ or $\vec{G}$) given by
Eq.~\eqref{eq:CoMcharge} differs from other definitions in the
literature~\cite{Blanchet:2024mnz, Compere:2019gft} by the factor of $P^{0}$
in the denominator. We denote these other PN definition of CoM charge by
$\vec{\mathcal{G}}$($ \equiv \vec{K} + u \vec{P}$) later in the text.
With the charges established we need to identify the BMS frames that we want our
numerical waveforms to be in.  This means to fix the rotations, proper
supertranslations, and finally, the subject of this work: translations and boosts.
The rotation freedom can be fixed during the inspiral
by computing an angular velocity vector $\vec{\Omega}$ that
keeps a waveform as constant as possible in the corotating frame.\footnote{%
This definition of $\vec{\Omega}$ coincides with the condition that waveform modes
from quasicircular binaries satisfy an approximate helical symmetry generated by
$\pd_{u}+(\vec{\Omega}\times\vec{x})\cdot\vec{\pd}$ as described
in~\cite{Khairnar:2024rzs}.} %
Such a
definition was built in~\cite{Boyle:2013nka,OShaughnessy:2011pmr}
\begin{align}
\label{eq:angularvelocity}
\vec{\Omega}(u)=-\langle\vec{L}\vec{L}\rangle^{-1} \cdot \langle\vec{L}\partial_{t}\rangle,
\end{align}
using the infinitesimal generators of rotations $\vec{L}$,
where the vector and matrix have components
\begin{subequations}
\begin{align}
\langle\vec{L}\partial_{t}\rangle^{a} & \equiv \sum\limits_{\ell,m,m'}\Im \left[\bar{f}_{\ell,m'}\langle\ell,m'|L^{a}|\ell,m\rangle\dot{f}_{\ell,m}\right],\\
\langle\vec{L}\vec{L}\rangle^{ab} & \equiv\sum\limits_{\ell,m,m'}\bar{f}_{\ell,m'}\langle\ell,m'|L^{(a}L^{b)}|\ell,m\rangle f_{\ell,m}.
\end{align}
\end{subequations}
Here $f(u, \t, \p)$ is an asymptotic waveform, e.g., the strain or news.
The angular velocity $\vec{\Omega}(u)$ can be computed at each instant $u$. To
fix a single global rotation, the \texttt{scri}
package~\cite{boyle_2025_15693419} effectively averages over a time window,
finding the average $\vec{\Omega}$ that best aligns the numerical timeseries
$\vec{\Omega}_{\text{NR}}$ with an analytical target
(e.g. $\vec{\Omega}_{\text{PN}}$ or simply the $z$ axis).
This leaves a
residual $U(1)$ freedom which can be fixed by, e.g., rotating about the
$\vec{\Omega}$ axis so that the $h_{(2,1)}$ mode is real and positive at some
time.

For fixing the supertranslation freedom, the Moreschi supermomentum is a
convenient choice for the supertranslation charge because it transforms in a
simple way under supertranslations, and vanishes in non-radiative spacetimes.
The Moreschi supermomentum is defined as~\cite{Moreschi:1988pc}
\begin{subequations}
\begin{align}
\label{eq:Moreschisupermomentum}
\Psi_{\text{M}}(u,\theta,\phi) &\equiv \Psi_{2}+\sigma\dot{\bar{\sigma}}+\eth^{2}\bar{\sigma}, \\
&=\int_{-\infty}^{u}|\dot{\sigma}|^{2}du-M_{\text{ADM}} \nonumber ,\\
&=\mathcal{E}-M_{\text{ADM}}.
\end{align}
\end{subequations}

For the case of compact binary systems
there are several related choices of BMS frame, all coming from the concept of a
``superrest'' frame.  A superrest frame is defined at some instant $u=u_{0}$ by
the vanishing of the $\ell\ge 1$ modes of the Moreschi supermomentum.
There are three convenient choices for $u_{0}$: (i) some slice during the
numerical evolution; (ii) the limit to arbitrarily late times, $u_{0}\to
+\infty$, and (iii) the limit to arbitrarily early times, $u_{0}\to -\infty$. %
Choice (i) is the simplest to implement. %
Choice (ii) is adapted to studying the remnant black hole. %
Choice (iii) agrees with assumptions of post-Newtonian waveform calculations, so
we call it the PNBMS frame here.
For the purpose of this work we are interested only in the PNBMS frame.

Thus, we need the expression of the BMS charges in the PNBMS frame.
The PN expressions of Moreschi supermomentum for the cases of
nonspinning and spinning binaries without eccentricity were computed in
\cite{Mitman:2022kwt}. Thus, the $\ell \geq 2$ supertranslations are found by
mapping $\Psi_M$ to the prediction coming from PN.

In Ref.~\cite{Mitman:2022kwt}, the space translations and the boost were found
by fitting the center-of-mass charge $\vec{G}$ to a linear function of time.
The slope and the intercept of the fit would give an estimate for the boost and
translation parameters. Essentially, they found a translation and boost that map
$\vec{G}$ to take an average of zero within some fitting window.

In this work we improve the center-of-mass frame fixing with a new procedure for
determining the boost and translation parameters. In the previous implementation
of the frame-fixing method, only the linear drift (seen in Fig.~\ref{fig:com
  charge}) was considered, while the oscillations were not accounted for. Thus,
the boost and translation parameters obtained in that way were more sensitive to
the location and size of the window. In the next section we present analytical
results derived from PN theory to reduce this parameter sensitivity.

\section{Modeling COM charge in PN theory}
\label{sec: frame fixing using PN}

In order to derive the PN expression for the center-of-mass charge, we use the
center-of-mass balance law derived by Comp\`ere et al.~\cite{Compere:2019gft}.
Starting from their expression in terms of the radiative multipole moments given
in their Eq.~(4.14b), we obtain the expression of center-of-mass flux in terms
of the source multipole moments to lowest PN order
\begin{align}
\dot{ \vec{\mathcal{G}}} &= \vec{P} + \dot{\vec{K}} + u \dot{\vec{P}} \\
\dot{ \vec{\mathcal{G}}} &= \vec{P} - \frac{G}{c^7} \left[ \frac{1}{21} \left( I^{(3)}_{jk} I^{(3)}_{ijk} - I^{(2)}_{jk} I^{(4)}_{ijk} \right) \right] + \mathcal{O}(c^{-9}). \label{eq:Compere-com-flux}
\end{align}
In contrast the definition of center-of-mass flux used by Blanchet et
al.~\cite{Blanchet:2018yqa,Blanchet:2024loi,Bernard:2017ktp} differs from the
previous equation by a total time derivative, and it gives a different result
for the CoM charge after using the equations of motion. We are using this
definition of the center-of-mass for the PN computation because it is identical
to the one implemented in the \texttt{scri} package~\cite{boyle_2025_15693419}
other than the factor of $P^0$.

For the case of quasicircular nonprecessing binaries, the linear momentum
charge at leading PN is given by~\cite{Wiseman:1992dv,Mishra:2011qz}
\begin{align}
\vec{P} = - \frac{464}{105} \frac{G^4 m^5}{c^7 r^4} \delta \, \nu^2 \hat{n} + \mathcal{O} \left(\frac{1}{c^{9}}\right).
\end{align}
We used the equations of motion from Blanchet's
review~\cite{Blanchet:2024mnz} to derive the contribution from the remaining
terms in Eq.~\eqref{eq:Compere-com-flux} at leading PN order. Therefore, we get
the total center-of-mass flux as
\begin{align}
\dot{ \vec{\mathcal{G}}} &=  \frac{1142}{105} \frac{G^4 m^5}{c^7 r^4} \delta \, \nu^2 \hat{n}  + \mathcal{O} \left(\frac{1}{c^{9}}\right).
\end{align}
We integrate this expression to obtain the center-of-mass charge at leading PN
order. The leading order result is given by
\begin{subequations}
\begin{align}
\vec{\mathcal{G}} &=  - \frac{1142}{105} \frac{G^4 m^5}{c^7 r^4 \Omega} \delta \, \nu^2 \hat{\lambda} + \mathcal{O} \left(\frac{1}{c^{9}}\right), \\
&= - \frac{1142}{105} \frac{G m^2}{c^2} x^{5/2} \sqrt{1-4\nu} \, \nu^2 \hat{\lambda} + \mathcal{O} \left(x^3\right)
\,.
\end{align}\label{eq:CoM in PN frame}%
\end{subequations}%
In these equations we make use of several PN quantities: our binary has total mass
$m=m_{1}+m_{2}$, the fractional mass difference
$\delta = (m_{1}-m_{2})/m = \sqrt{1 - 4 \nu}$, the symmetric mass ratio
$\nu=m_{1}m_{2}/(m_{1}+m_{2})^{2}$, the separation $r$, the frequency $\Omega$,
the PN order counting parameter $x$, and $\hat{\lambda}$ is the unit vector
in the direction of motion of the reduced mass.  In the quasicircular case, $x$
is defined as $x\equiv(Gm\Omega/c^{3})^{2/3}$.  The frequency $\Omega$ is the
``orbital'' frequency but as measured by asymptotic observers, for example
$\Omega = \Omega_{2,2}/2$ where $\Omega_{2,2}$ is the frequency of the (2,2)
mode of the waveform.\footnote{%
See~\cite{Trestini:2025nzr} for an illuminating discussion on the difference in
calculations with $\Omega$ vs. the orbital frequency measured by near-zone
observers, $\omega$.  This also imprints on the difference in PN expansions with
respect to $x$ or with respect to $y\equiv(Gm\omega/c^{3})^{{2/3}}$, but $x$ and
$y$ differ only at 4PN and higher.  Note that~\cite{Trestini:2025nzr} uses
$\Omega_{22}$ to represent \emph{half} of the frequency of the (2,2) mode of the
waveform.}

$\vec{\mathcal{G}}$ is in the direction of $\hat{\lambda}$ up to 2PN order. At
relative 2.5PN order we expect a contribution in the direction of $\hat{n}$, the
unit vector pointing from the center of mass to the position of the reduced
mass. We use this fact while numerically fitting the center-of-mass charge
later.

\figfvsnu

The factor $\nu^2 \sqrt{1-4\nu}$ is
crucial during the numerical computation of the CoM charge. It vanishes
at the equal-mass point $\nu = 1/4$, where moreover its derivative
is undefined. If there are small errors in the numerical
estimate of $\nu$ from the metadata, it can result in large differences
between the numerical computation of $\vec{G}$ and the analytical expression
from Eq.~\eqref{eq:CoM in PN frame}. We show this behavior of the prefactor in
Fig.~\ref{fig:fvsnuplot}. Thus, we avoid using equal mass ratio binaries for our
analysis.  Pushing the calculation in Eq.~\eqref{eq:CoM in PN frame} to
higher PN order could potentially alleviate this sensitivity to $\nu$.

The NR simulations are in frames which are boosted and translated with respect to
the PN CoM frame. Thus, we need to know how $\vec{G}$ transforms under
translations and (small) boosts, to be able to perform fits to NR data.
We start with the expression of center-of-mass charge
\begin{align}
    \vec{G} = \frac{1}{P^0}(\vec{K} + u \vec{P}) = \frac{\vec{\mathcal{G}}}{P^0}. \label{eq:com-to-boost-relation}
\end{align}
For a boost parameter $\vec{\beta}$,\footnote{%
This boost vector $\vec{\beta}$ should not be confused with the metric
function $\beta$ defined in Sec.~\ref{sec: effect of BMS}.} the individual terms will transform under
a small boost as
\begin{subequations}
\begin{align}
\vec{K}' &= \vec{K} - \vec{\beta} \times \vec{J} + \mathcal{O}(\beta^2),
\label{eq:boost_charge_transform}\\
\vec{P}' &= \vec{P} - P^0 \vec{\beta} + \mathcal{O}(\beta^2), \label{eq:angmom_charge_transform}\\
P^{0'} &= P^0 - \vec{\beta} \cdot \vec{P}+ \mathcal{O}(\beta^2), \label{eq:energy_charge_transform}\\
u' &= u(1 - \vec{\beta} \cdot \hat{n}) + \mathcal{O}(\beta^2) \,.
\end{align}
\end{subequations}
Thus, the transformation of $\vec{G}$ under a small boost is given by
\begin{align}
\vec{G}' = & \frac{1}{P^{0}} \Big\{\vec{K} - \vec{\beta} \times \vec{J} + u 
\left( \vec{P} - P^0 \vec{\beta} - \vec{\beta} \cdot \hat{n}\vec{P} \right) \nonumber \\
 & - \frac{ \vec{\beta} \cdot \vec{P} }{P^0} \vec{K} -  \frac{\vec{\beta} \cdot 
\vec{P}}{P^0} u \vec{P}\Big\} + \mathcal{O}(\beta^2), \label{eq:boosted G}
\end{align}
where we have neglected $\mathcal{O}(\beta^2)$ terms.
Since we are using PN results, all the charges have expansions in terms
of the PN order-counting parameter $x$. We use the PN expression of
BMS charges for quasicircular nonprecessing systems presented
in~\cite{Blanchet:2018yqa, Blanchet:2024mnz}, (we only need up to 3PN order for
our level of approximation)
\begin{subequations}
\begin{align}
P^0 &= E =  \, m - \frac{m \nu x}{2} \Bigg\{ 1 + \left( - \f{3}{4} - \f{\nu}{12} \right) x \nn \\
& + \left(- \f{27}{8} + \f{19}{8}\nu - \f{\nu^2}{24} \right)x^2 + \mathcal{O}(x^3) \Bigg\} \,, \\
J &= \f{\nu m^2}{x^{1/2}} \Bigg\{ 1 + \left( \f{3}{2} + \f{\nu}{6} \right) x 
+ \left(\f{27}{8} - \f{19}{8}\nu + \f{\nu^2}{24} \right)x^2 \nn \\
& + \left(\f{135}{16} + \left[ \f{-6889}{144} + \f{41}{24} \pi^2 \right] \nu + \f{31}{24} \nu^2 + \f{7}{1296} \nu^3 \right) x^3 \nn \nn \\
&  + \mathcal{O}(x^4) \Bigg\} \,.
\end{align}	\label{eq:E and J series}
\end{subequations}

Upon reexpanding Eq.~\eqref{eq:boosted G} using the PN series in
Eq.~\eqref{eq:CoM in PN frame} and Eq.~\eqref{eq:E and J series}, the terms that will have the dominant effect will be
\begin{align}
\vec{G}' &= \frac{1}{P^0} (\vec{\mathcal{G}} - \vec{\beta} \times \vec{J} - u P^0 \vec{\beta}) + \mathcal{O}(x,\beta^2), 
\end{align}
where we have used $\vec{\mathcal{G}} = \vec{K} + u \vec{P}$ to combine terms.
As mentioned earlier we expect higher PN order corrections to $\vec{\mathcal{G}}$ in the
direction of $\hat{n}$, and errors in the estimation of $\nu$ obtained from the
metadata. Therefore, we introduce nuisance parameters
$\alpha_{i}$ (for $i=1,2$) to account for this difference in amplitude and
direction bewteen NR and PN waveforms. Finally, we add parameters $\vec{\Delta}$
for the translation of the origin of the CoM at time $u=0$.
Thus, our fitting function is
\begin{align}
\vec{G}' = & \frac{1}{P^0} \left((\alpha_1 \hat{\lambda} + \alpha_2 \hat{n})|\vec{\mathcal{G}}|    - \vec{\beta} \times \vec{J} - u P^0 \vec{\beta}\right) \nonumber \\
 & + \vec{\Delta} + \mathcal{O}(x^3,\beta^2) \,,
\label{eq:Boosted CoM charge}
\end{align}
where $|\vec{\mathcal{G}}|$ is the magnitude of the leading order term in Eq.~\eqref{eq:CoM in PN frame}.

\section{Sensitivity analysis}
\label{sec:sensitivity-analysis}

\tabSimulations

To compare the previous implementation to our new approach, we quantified the
sensitivity of boost and translation fit parameters---on the size and location
of the mapping window---using NR waveforms from the SXS catalog. The size of the
fitting window governs the number of orbital cycles that are used during
the fitting process, while the location controls the effects from the junk in
early inspiral and higher order PN effects in late inspiral.

We selected a set of 20 quasicircular nonprecessing systems with eccentricity
$e<10^{-4}$ \cite{Scheel:2025jct,Boyle:2019kee} for our numerical analysis,
listed in Table~\ref{tab:simulations}. These simulations have mass ratios
$1.2<q<9$. For cases very close to $q=1$, we noticed
large differences between the analytical and numerical center of
mass charge $\vec{G}$ (as discussed in the previous section).
Thus we restricted our analysis to $q>1.2$. For every simulation, we extracted
the gravitational wave strain and Weyl scalars at future null infinity using the
public SpECTRE code's CCE
module~\cite{Moxon:2020gha,Moxon:2021gbv,spectrecode}. This data will be made
publicly available in the upcoming SXS Collaboration's CCE catalog.

\figWindowChoices
\figCombinedBoostTransComp

Our goal is to inspect the sensitivity of the boost and translation fit
parameters by varying the window size and its location. There are 3 ways to
place a window over the inspiral and vary its size, as shown in
Fig.~\ref{fig:window-choices}: (1) fixing the start, (2) fixing the center, and
(3) fixing the end of the window.
We consider the inspiral to take place between a simulation's
$\texttt{reference\_time}$ (available in the SXS catalog
metadata~\cite{Scheel:2025jct}) and the time of the maximum of the $L^{2}$ norm
of $h$ across the entire 2-sphere, available through the function
$\texttt{max\_norm\_time}$ in the \texttt{scri}
package~\cite{boyle_2025_15693419}.
For the cases of fixing the starting and ending side, we fixed them at $10\%$
(of the inspiral length) away from $\texttt{reference\_time}$ and the
$\texttt{max\_norm\_time}$
respectively. Then we varied the other end uniformly thus increasing the size of
the window. For case (2) we found the center of the inspiral using the
$\texttt{reference\_time}$ and $\texttt{max\_norm\_time}$, and increased the
window size in both directions. Thus, the range of window size increases
from approximately 500M to 4000M for a typical simulation. These choices reduce
the contamination from junk radiation and late-merger effects where PN
approximations break down, though some residual effects may remain.

We use the asymptotic Bondi data
evolved with CCE for our analysis. The center-of-mass charge
$\vec{G}$ can be obtained numerically from the asymptotic data as discussed in
Section \ref{sec: frame fixing using charges}. The boosted center-of-mass charge
is computed numerically using Eq.~\eqref{eq:Boosted CoM charge},
Eq.~\eqref{eq:CoM in PN frame}, and Eq.~\eqref{eq:E and J series}. The unit vectors
parameterizing the orbit are numerically defined as
\begin{subequations}
\begin{align}
	\hat{n} &= \cos{\psi} \, \hat{e}_x + \sin{\psi} \, \hat{e}_y, \\
	\hat{\lambda} &= - \sin{\psi} \, \hat{e}_x + \cos{\psi} \, \hat{e}_y,
\end{align}
\end{subequations}
where $\hat{e}_x$ and $\hat{e}_y$ are Cartesian basis vectors defined at some
instant when the reduced mass is along the positive $x$ axis, with velocity
along the positive $y$ axis.
Here $\psi$ is the gravitational wave phase
derived from the $h_{2,1}$ mode of the waveform.
Because of the difference in conventions between PN theory and NR, we evaluate
it as
\begin{align}
\psi = - \arg(-h_{2,1}^{\text{NR}}) + \f{\pi}{2}.
\end{align}
Concretely, SpEC defines the metric perturbation as
$h^{\text{NR}}_{ab} = g_{ab}-\eta_{ab}$, while PN theory~\cite{Blanchet:2024mnz}
defines the metric perturbation as
$h_{\text{PN}}^{ab} = \sqrt{|g|}g^{ab}-\eta^{ab}$, resulting in
$h_{\ell,m}^{\text{NR}} = -h_{\ell,m}^{\text{PN}}+\mathcal{O}(h^{2})$.  The $\pi/2$ phase difference is due
to the leading order complex phase of $h_{2,1}$ mode from PN theory,
\begin{align}
h_{2,1}^{\text{PN}} = \f{2 G \mu x}{R c^2}
\sqrt{\f{16 \pi}{5}} \left( \f{1}{3} i \, \delta \, x^{1/2} + \mathcal{O}(x^{3/2}) \right) e^{- i \psi}.
\end{align}
The PN parameter $x$ is obtained numerically from the magnitude of angular velocity
evaluated from the strain as defined in Eq.~\eqref{eq:angularvelocity}.
The numerically-obtained center-of-mass charge vector is fit to the boosted
charge expression given in Eq.~\eqref{eq:Boosted CoM charge} using
least squares. Figure~\ref{fig:com charge} shows the PN fit (orange
dashed curve) plotted over the numerical $\vec{G}$ (blue curve). Clearly, the PN
expression fits the data very well. The analytical result models the
oscillations as well as the linear trend. Therefore, we use the PN expression to
estimate the boost and translation required to map our system to the CoM frame.

We compare the robustness of the fit parameters using our new method
against the previous method
in Fig.~\ref{fig:sensitivity plot}. As expected the boost $\vec{\beta}$ and
translation $\vec{\Delta}$ parameters are less sensitive when the fit is
performed using the PN method. The fit values also converge with smaller window
size---as small as 1500M---for the PN-based method as compared to the previous
linear fit. Thus, we recommend users to consider window sizes of at least 1500M
during the frame fixing procedure. Our analysis is confined to the region of
quasicircular, nonprecessing parameter space. Hence, our PN results indicate
that there is no motion of the center of mass in the direction perpendicular to
the orbital plane. Therefore, we continue modeling the $z$ component of the
center-of-mass charge with a linear function of time as done previously.

In order to quantify the improvement, we computed the variance of the fit
parameters ($\beta^x$, $\beta^y$, $\Delta^x$, and $\Delta^y$) for changing
window size for each simulation. We compare the performance of the two methods by
calculating the ratio of variance obtained for each parameter (e.g.,
$\sigma^2_{\text{linear}}(\beta^x)/\sigma^2_{\text{PN}}(\beta^x)$). Given a
simulation, this ratio will be closer to 1 when the sensitivity of the fit from
two methods are comparable. Conversly, the ratio will be larger than 1 when the PN-based method is more robust than the previous linear method. We present
the median value of this ratio across 20 simulations for each parameter, and for
the 3 different cases of placing the window in
Table~\ref{tab:fit-median-improvement}.

\tabfitmedianimprovement

Our new method outperforms the previous
method, improving the robustness of fit parameters by a factor of $\sim 25$
for $\beta^x$ and a factor of $\sim 20$ for $\Delta^{x}$ when the center of the window was fixed.
This is the case with the best improvement. Contrary to our expectation, we
observed the least improvement for the case when the start of the window was
fixed. We speculate that this is due to the effects of junk radiation present in
the center-of-mass charge long after the relaxation time. The treatement of junk
radiation is beyond the scope of this work, so we do not address it here. Also,
we notice a decent improvement in robustness for the case of fixing the end
side. The leading PN order result gives a satisfactory fit in the late
inspiral regime as well. The center fixed case is less affected by the
combination of junk radiation and higher order PN effects, hence it exhibits the
best improvement.

\figAlphas

Besides the above analysis, we also inspected the fit values for the nuisance
parameters. For a fixed window size of 1500M placed at the center, we
plot the values of $(\alpha_1,\alpha_{2})$ for all the simulations in
Fig.~\ref{fig:alphas}. From the PN prediction, we expect these
coefficients to take values close to $(\alpha_{1},\alpha_{2}) = (1,0)$ (green
cross), with reasonable deviations due to higher order PN effects. We
couldn't address the behavior of the cases that are deviating significantly
from the expected value, and we leave it for
future work. We have examined that this plot does not depend significantly on
the size of the 
window or its location. We placed the window at least $15\%$ (of the inspiral
length) away from the \texttt{reference\_time} for producing the results of
Fig.~\ref{fig:sensitivity plot} and Fig.~\ref{fig:alphas}. This ensured that the
effects from junk radiation were minimized for the entire analysis. Therefore,
we do not expect junk radiation to be the cause of the deviations in the
nuisance parameters.

Finally, we incorporate this method into the frame
fixing routine defined in the python package \texttt{scri}
\cite{boyle_2025_15693419,Boyle:2013nka,Boyle:2014ioa,Boyle:2015nqa}. Frame
fixing is implemented by the \texttt{map\_to\_superrest\_frame} function of
\texttt{scri}. We extended this function to accept user-defined fitting functions
along with fit parameters and initial guesses. The resulting fit
parameters are passed to the existing BMS
transformation pipeline. The structure of this function is described in
detail in the documentation of the \texttt{scri} package. We also provide an
example script in the documentation, implementing the function used in this
work, which we recommend for direct use.

\section{Discussion and conclusions}
\label{sec:disc-concl}

Gravitational waveforms extracted from an NR simulation are always in some
arbitrary BMS frame. This is due to gauge choices made during the setup and
evolution of the NR simulation. SpEC simulations, when extracted via
extrapolation or evolved with CCE, are no exception. Because of these gauge
choices, the power of the dominant mode leaks into higher order modes. These
features are not expected from analytical models, as all of them
are constructed in a center-of-mass gauge.
These gauge effects needs to be fixed to directly compare numerical and
analytical waveforms. This also simplifies numerical waveforms by gauging away
the oscillations seen in Fig.~\ref{fig:amplitudefig}.

For this work, we focused on waveforms evolved using the CCE method.
We presented an improvement in the procedure of fixing the PNBMS frame of CCE
waveforms from the SXS catalog. Previous studies have demonstrated the use of
BMS charges to fix the BMS frame. In particular, the supertranslations were
fixed using the PN Moreschi supermomentum charge, rotations were fixed using
the angular velocity derived from waveforms,
and boosts along with translations were fixed using the
CoM charge. These charges are computed numerically from the Weyl scalars, and
thus all the BMS freedom was fixed.

In this study, we presented an improvement to the method of finding the boost
and translations from the CoM charge, which is a subset of the full
frame fixing routine. The numerical CoM charge exhibits both a linear drift,
indicating that the binary is not in the CoM frame, and
physical oscillations on the orbital timescales. In earlier approaches, only a
linear fit to the CoM charge was used to determine
the boost and translation from the slope and intercept. Here, we instead modeled
both the linear drift and the oscillatory behavior using an analytical
expression of the CoM charge derived from PN theory. Specifically, we employed
the center-of-mass balance law, and the PN equations of motion to derive the CoM
charge for a quasicircular, nonprecessing binary at leading PN order. We also
derived an analytical fitting function, which we used to fit the numerical
boosted CoM charge.

We found that our PN-inspired approach gives a more robust set of boost and
translation parameters to fix the frame. Our fit parameters are found to be less
sensitive to the size and location of the window over which the frame fixing
routine is performed. The largest improvement is by a factor of $\sim 25$ for
the boost parameter, and $\sim 20$ for the translation parameter when the
mapping window is placed at the center of the inspiral. We recommend users
to use a mapping window of size at least $\sim 1500M$, and position it near the
center of the inspiral while fixing the PNBMS frame. This choice minimizes
the errors from junk radiation and neglecting higher-PN terms.
We performed further analysis to quantify the effect of our PN-inspired
method on waveform modes. We found the mismatches to be small — which is
consistent with the small magnitude of the boost and translation parameter
differences between the two methods. The primary impact is not on the waveform
modes, but rather on the robustness and consistency of the fitting procedure
itself, which now relies on analytical PN results.

In future work, we would like to extend this approach for eccentric and precessing
cases, deriving the necessary PN results. We also want to extend the PN
computation of CoM charge to higher PN order for quasicircular nonprecessing
systems. Additionally, we would like to derive the analytical result for boosted
CoM charge using charge integrals or through the transformations of BMS charges.

\acknowledgments
The authors would like to thank David Trestini and Neev Khera for useful
discussions regarding computation of CoM charge. This work was supported by NSF
CAREER Award PHY–2047382 and a Sloan Foundation Research Fellowship. K.M.\ is supported by NASA through the NASA Hubble Fellowship grant \#
HST-HF2-51562.001-A awarded by the Space Telescope Science Institute,
which is operated by the Association of Universities for Research in
Astronomy, Incorporated, under NASA contract
NAS5-26555. This material is based upon work supported by the National Science
Foundation under Grants No.~PHY-2309211; No.~PHY-2309231; No.~OAC-2513339 at
Caltech; and NASA award No.~80NSSC26K0340, and No.~PHY-2407742; No.~PHY-2207342;
No.~OAC-2513338; and NASA award No.~80NSSC26K0340 at Cornell. Any opinions,
findings, and conclusions or recommendations expressed in this material are
those of the author(s) and do not necessarily reflect the views of the National
Science Foundation or NASA. This work was supported by the Sherman Fairchild
Foundation at Caltech and Cornell. Some numerical computations were performed on
the Maple cluster at the Mississippi Center for Supercomputing Research (MCSR)
at the University of Mississippi.

\appendix

\bibliographystyle{JHEP}
\bibliography{notes-biblio}

\end{document}